\documentclass[12pt]{article}
\usepackage{graphicx}

\newcommand\pubnumber{}
\newcommand\pubdate{\today}

\def\york{Department of Math and Statistics\\
York University, 4700 Keele St., Toronto ON CANADA M3J 1P3\\
and\\
CSSM, University of Adelaide, Adelaide SA 5005 Australia}
\def\support{\footnote{Work supported by the Natural Sciences and
Engineering Research Council of Canada}}

\def\Title#1{\begin{center} {\Large #1 } \end{center}}
\def\Author#1{\begin{center}{ \sc #1} \end{center}}
\def\Address#1{\begin{center}{ \it #1} \end{center}}

\newcommand\pubblock{\rightline{\begin{tabular}{l} \pubnumber\\
         \pubdate  \end{tabular}}}
\newenvironment{Abstract}{\begin{quotation}  }{\end{quotation}}
\newenvironment{Presented}{\begin{quotation} \begin{center} 
             PRESENTED AT\end{center}\bigskip 
      \begin{center}\begin{large}}{\end{large}\end{center} \end{quotation}}

\def\beq{\begin{equation}}
\def\eeq#1{\label{#1}\end{equation}}
\def\eeqn{\end{equation}}

\def\beqa{\begin{eqnarray}}
\def\eeqa#1{\label{#1}\end{eqnarray}}
\def\eeqan{\end{eqnarray}}

\let\bar=\overbar

\def\Dslash{\not{\hbox{\kern-4pt $D$}}}
\def\dslash{\not{\hbox{\kern-2pt $\del$}}}

\def\msb{{\bar{\ssstyle M \kern -1pt S}}}

\begin{document}
\begin{titlepage}
\pubblock

\vfill
\Title{Hadronic $\tau$ Decay Based Determinations of $\vert V_{us}\vert$}
\vfill
\Author{ Kim Maltman\support}
\Address{\york}
\vfill
\begin{Abstract}
I review sum rule determinations of $\vert V_{us}\vert$ employing hadronic
$\tau$ decay data, taking into account recent HFAG updates of exclusive
$\tau$ branching fractions and paying special attention to the impact of 
the slow convergence of the relevant integrated $D=2$ OPE series and
the potential role of contributions of as-yet-unmeasured higher 
multiplicity modes to the strange inclusive spectral distribution. 
In addition to conventional flavor-breaking sum rule determinations, 
information obtainable from mixed $\tau$-electroproduction sum rules having
much reduced OPE uncertainties, and from sum rules based on
the inclusive strange decay distribution alone, is also considered.
Earlier discrepancies with the expectations of 3-family unitarity
are found to be reduced, both the switch to $D=2$ OPE treatments
favored by self-consistency tests and the 
increase in the strange branching fractions playing a role in
this reduction.
\end{Abstract}
\vfill
\begin{Presented}
CKM2010, the 6th International Workshop on the CKM Unitarity
Triangle, University of Warwick, UK, 6--10 September 2010
\end{Presented}
\vfill
\end{titlepage}
\def\thefootnote{\fnsymbol{footnote}}
\setcounter{footnote}{0}

\section{Introduction}

Recent determinations of $\vert V_{us}\vert$ using flavor-breaking 
(FB) hadronic $\tau$ decay sum 
rules~\cite{gamizetal,kmcwvus,gamizetalnew,hfag10}
yield results $\sim 3\sigma$ low compared to both 3-family
unitarity expectations, and those from
$K_{\mu 3}$ and $K_{\mu 2}$ analyses~\cite{htvud,kell3andKratiosvus}.
The $\tau$ determinations employ finite energy sum rules (FESRs)
which, for a kinematic-singularity-free correlator, $\Pi$, with spectral
function, $\rho$, take the form (valid for arbitrary $s_0$ and
analytic $w(s)$)
\begin{equation}
\int_0^{s_0}w(s) \rho(s)\, ds\, =\, -{\frac{1}{2\pi i}}\oint_{\vert
s\vert =s_0}w(s) \Pi (s)\, ds\ .
\label{basicfesr}
\end{equation}
$\vert V_{us}\vert$ is obtained by setting $\Pi =\Delta\Pi_\tau \, \equiv\,
\left[ \Pi_{V+A;ud}^{(0+1)}\, -\, \Pi_{V+A;us}^{(0+1)}\right]$,
with $\Pi^{(J)}_{V/A;ij}(s)$ the spin $J=0,1$ components
of the flavor $ij$, vector (V) or axial vector (A) current two-point
functions.
For large enough $s_0$, the OPE can be used on the RHS, while 
for $s_0\leq m_\tau^2$, the $\rho^{(J)}_{V/A;ij}$ needed on
the LHS are related to the inclusive differential
distributions, $dR_{V/A;ij}/ds$, with
$R_{V/A;ij}\, \equiv\, \Gamma [\tau^- \rightarrow \nu_\tau
\, {\rm hadrons}_{V/A;ij}\, (\gamma )]/ \Gamma [\tau^- \rightarrow
\nu_\tau e^- {\bar \nu}_e (\gamma)]$, by~\cite{tsai}
\begin{equation}
{\frac{dR_{V/A;ij}}{ds}}\, =\, {\frac{12\pi^2\vert V_{ij}\vert^2 S_{EW}}
{m_\tau^2}}\,
\left[ w_\tau (y_\tau ) \rho_{V/A;ij}^{(0+1)}(s)\ \
- w_L (y_\tau )\rho_{V/A;ij}^{(0)}(s) \right]
\label{basictaudecay}\end{equation}
with $y_\tau =s/m_\tau^2$, $w_\tau (y)=(1-y)^2(1+2y)$,
$w_L(y)=2y(1-y)^2$, $V_{ij}$ the flavor $ij$ CKM matrix element,
and $S_{EW}$ a short-distance electroweak correction.

The $J=0+1$ combination, $\Delta\Pi_\tau$, is employed due to
the extremely bad behavior of the integrated
$J=0$, $D=2$ OPE series~\cite{longprob}.
Fortunately, $J=0$ spectral contributions are dominated
by the accurately known $K$ and $\pi$ pole terms,
with residual continuum contributions
numerically negligible for $ij=ud$,
and determinable phenomenologically 
via dispersive~\cite{jop} and sum rule~\cite{mksps} analyses
for $ij=us$. Subtracting the $J=0$ contributions from $dR_{V+A;ij}/ds$,
one can evaluate the re-weighted $J=0+1$ 
integrals $R^w_{V+A;ij}(s_0)\equiv 12\pi^2 S_{EW} \vert V_{ij}\vert^2
\int_0^{s_0}{\frac{ds}{m_\tau^2}}\, w(s)\, \rho^{(0+1)}_{V+A;ij}(s)$
and FB differences
\begin{eqnarray}
&&\delta R^w_{V+A}(s_0) =
{\frac{R^w_{V+A;ud}(s_0)}{\vert V_{ud}\vert^2}}
 - {\frac{R^w_{V+A;us}(s_0)}{\vert V_{us}\vert^2}}
=12\pi^2 S_{EW} 
\int_0^{s_0}{\frac{ds}{m_\tau^2}}\, w(s)\Delta\rho_\tau (s)\ .
\end{eqnarray}
Taking $\vert V_{ud}\vert$ and any OPE parameters from other sources,
Eq.~(\ref{basicfesr}) then yields~\cite{gamizetal}
\begin{equation}
\vert V_{us}\vert \, =\, \sqrt{R^w_{V+A;us}(s_0)
/\left[{\frac{R^w_{V+A;ud}(s_0)}{\vert V_{ud}\vert^2}}
\, -\, \delta R^{w,OPE}_{V+A}(s_0)\right]}\ .
\label{tauvussolution}\end{equation}
The OPE contribution in Eq.~(\ref{tauvussolution})
is at the few-to-several-$\%$ level of the $ud$ spectral
integral term for weights used previously in the 
literature~\cite{gamizetal,kmcwvus,gamizetalnew}, making modest accuracy 
for $\delta R^{w,OPE}_{V+A}(s_0)$ sufficient for a high 
accuracy determination of $\vert V_{us}\vert$
{\footnote{As an example, 
removing entirely the OPE corrections from the recent HFAG 
$s_0=m_\tau^2$, $w=w_\tau$ determination, 
$\vert V_{us}\vert$ is shifted by only $\sim 3\%$, from 
$0.2174(23)$~\cite{hfag10} to $0.2108(19)$.}}.

Estimating the error on $\delta R^{w_\tau ,OPE}_{V+A}(s_0)$ is complicated 
by the slow convergence of the leading dimension $D=2$ OPE series,
$\left[\Delta \Pi_\tau \right]_{D=2}^{OPE}$. To four loops~\cite{bckd2ope}
\begin{equation}
\left[\Delta\Pi_\tau (Q^2)\right]^{OPE}_{D=2}\, =\, {\frac{3}{2\pi^2}}\,
{\frac{m_s(Q^2)}{Q^2}} \left[ 1 + {\frac{7}{3}} \bar{a}
+ 19.93 \bar{a}^2  + 208.75 \bar{a}^3
 + d_4 \bar{a}^4 + \cdots \right]\ \ \
\label{d2form}\end{equation}
with $\bar{a}=\alpha_s(Q^2)/\pi$, and $\alpha_s(Q^2)$ and $m_s(Q^2)$
the running coupling and strange quark mass in the $\overline{MS}$ scheme
{\footnote{We use the estimate $d_4\sim 2378$~\cite{bckd2ope}
for the as-yet-undetermined 5-loop coefficient $d_4$.}}.
Since $\bar{a}(m_\tau^2)\simeq 0.1$, convergence at the spacelike point on
$\vert s\vert = s_0$ is marginal at best and conventional error 
estimates may significantly underestimate 
the truncation uncertainty. Consistency checks are, however, possible.
Assuming both the data and OPE error estimates are 
reliable, $\vert V_{us}\vert$ 
should be independent of $s_0$ and $w(s)$. On the
OPE side, results obtained using $D=2$ truncation schemes
differing only at orders beyond the truncation order
should agree to within the truncation uncertainty estimate.
We consider three commonly used truncation schemes: 
the contour improved (CIPT) prescription, used with either 
the truncated expression for
$\left[ \Delta\Pi_\tau\right]^{OPE}_{D=2}$, or, after
partial integration, the correspondingly truncated Adler function series,
and the truncated fixed-order (FOPT) prescription.

\section{$\vert V_{us}\vert$ from various FESRs employing
$\tau$ decay data}

Results below are based on updated 2010 HFAG hadronic
and lepton-universality-constrained leptonic $\tau$ BFs~\cite{hfag10}, 
supplemented by SM $K_{\mu 2}$ and $\pi_{\mu 2}$ expectations for $B_K$
and $B_\pi$. The publicly available ALEPH $ud$ 
distribution~\cite{alephud05}, rescaled to reflect the resulting
normalizations $R_{V+A;us}=0.1623(28)$, $R_{V+A;ud}=3.467(9)$,
is used for $\rho_{V+A;ud}(s)$. Though improved exclusive $us$ BFs
are available from BaBar and Belle, a completed inclusive $us$ distribution
is not. The ALEPH inclusive $us$ distribution~\cite{alephus99}, however,
corresponds to exclusive BFs with significantly larger errors,
and, sometimes, significantly different central values~\cite{hfag10}.
Following Ref.~\cite{alephusrescaleidea},
we ``partially update'' $\rho_{V+A;us}(s)$,
rescaling the ALEPH distribution mode by mode 
with the ratio of new to old BFs. This procedure
works well when tested using BaBar
$\tau\rightarrow K^-\pi^+\pi^-\nu_\tau$ data~\cite{babarkppallchg}, 
but is likely less reliable for modes 
($K3\pi ,\, K4\pi ,\cdots$) estimated using Monte Carlo
rather than measured by ALEPH.
OPE input is specified in Ref.~\cite{kmtau2010}.

For $s_0=m_\tau^2$, $w=w_\tau$, the $ud$ and $us$ spectral integrals 
needed in the FB $\Delta\Pi_\tau$ FESR
are determined by the corresponding inclusive BFs.
Conventional last-term-retained$\oplus$residual-scale-dependence 
$D=2$ OPE truncation error estimates yield a combined 
theoretical uncertainty of $0.0005$ on $\vert V_{us}\vert$
in this case~\cite{gamizetalnew}.

\begin{figure}[htb]
\centering
\rotatebox{270}{\mbox{
  \begin{minipage}[t]{2.3in}
\includegraphics[width=2.2in]{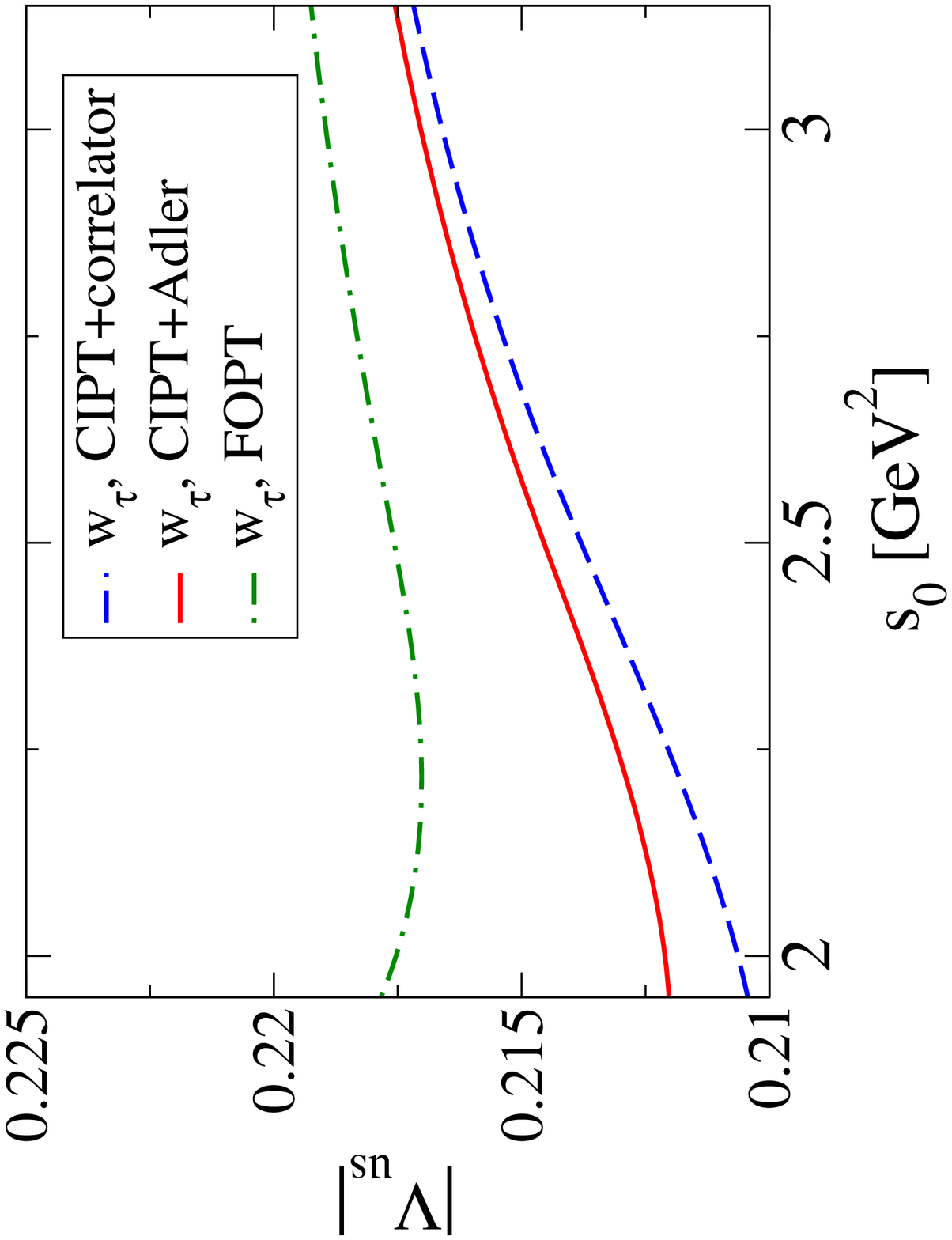}
  \end{minipage}}}\ \ 
\rotatebox{270}{\mbox{
  \begin{minipage}[t]{2.3in}
\includegraphics[width=2.2in]{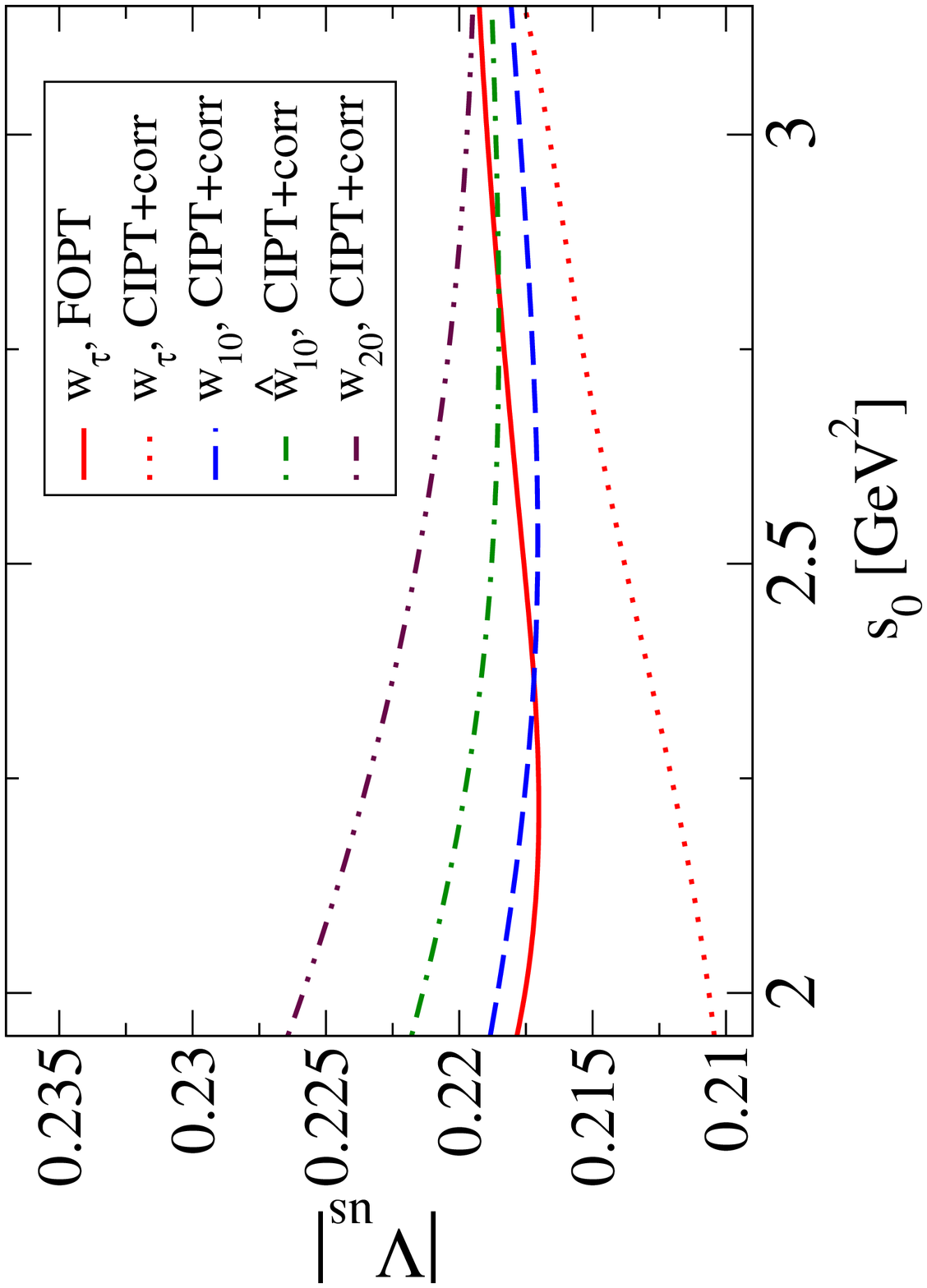}
  \end{minipage}}}\ \ 

\caption{$\vert V_{us}\vert$ vs. $s_0$ for (i) Left panel: the FB $w_\tau$ 
FESR, using the three prescriptions for the $D=2$ OPE series and (ii)
Right panel: the FB $w_{10}$, $\hat{w}_{10}$ and $w_{20}$ FESRs,
using the CIPT+correlator prescription, with FB $w_\tau$ results shown
for comparison.}
\label{udusfbfesrs}
\end{figure}

The left panel of Fig.~\ref{udusfbfesrs} shows $\vert V_{us}\vert$
versus $s_0$ for each of the three
prescriptions for the $w_\tau$-weighted $D=2$ OPE series.
The two CIPT treatments give similar results,
but show poor $s_0$-stability. The FOPT prescription
yields significantly improved, though not perfect, $s_0$-stability.
For all $s_0$, the FOPT-CIPT difference is significantly greater
than the nominally estimated $0.0005$ theoretical error.
The integrated $D=2$ series is also better behaved for FOPT.
The FOPT version of $\delta R^{w_\tau,OPE}_{V+A}(m_\tau^2)$ is
a factor of $\sim 2$ larger than either of the two CIPT versions,
suggesting that the integrated $D=2$ convergence is indeed slow,
and the resulting truncation uncertainty large. The
$s_0=m_\tau^2$ version of the better behaved FOPT prescription
yields 
\begin{equation}
\vert V_{us}\vert = 0.2193(3)_{ud}(19)_{us}(19)_{th}\ ,
\end{equation}
$\sim 2.3\sigma$ below 3-family unitarity expectations,
the theory error reflecting the sizeable $D=2$ FOPT-CIPT difference.
The right panel of Fig.~\ref{udusfbfesrs} compares 
the results from FB FESRs corresponding to three additional weights, 
$w_{10}$, $\hat{w}_{10}$, and $w_{20}$, 
constructed in Ref.~\cite{km00} to improve
convergence of the integrated CIPT $D=2$ series, with those
of the $w_\tau$ case.
Improved $s_0$-stability is observed, together with a reduced 
weight-choice dependence.
For $\hat{w}_{10}$ (which shows the best $s_0$-stability),
$\vert V_{us}\vert = 0.2188$ at $s_0=m_\tau^2$.
In the absence of a new version of the inclusive $us$ distribution, 
the experimental error has to be based on
the 1999 ALEPH $us$ covariances, and is $0.0033$.

\begin{figure}[htb]
\centering
\rotatebox{270}{\mbox{
  \begin{minipage}[t]{2.3in}
\includegraphics[width=2.2in]{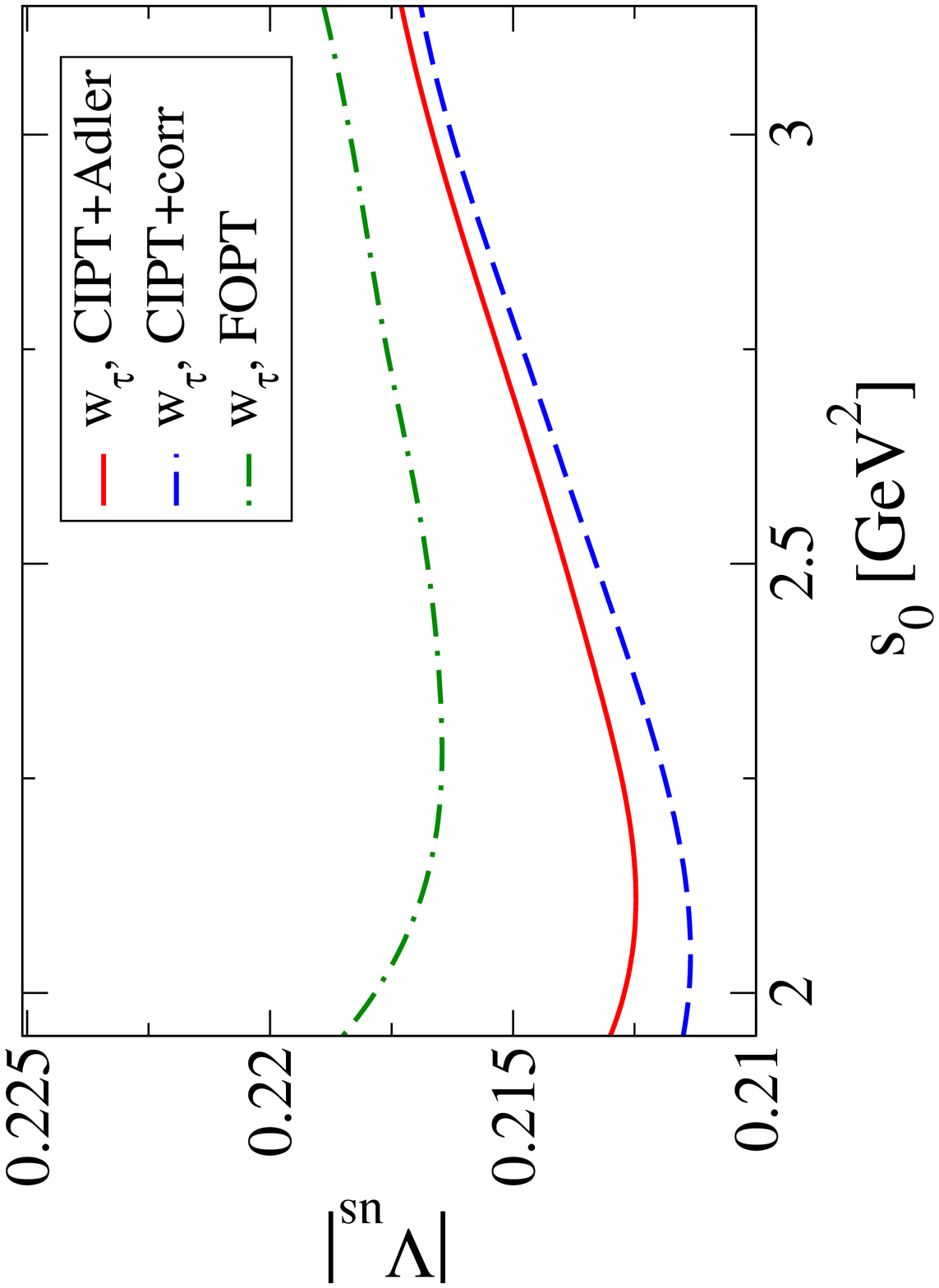}
  \end{minipage}}}\ \ 
\rotatebox{270}{\mbox{
  \begin{minipage}[t]{2.3in}
\includegraphics[width=2.2in]{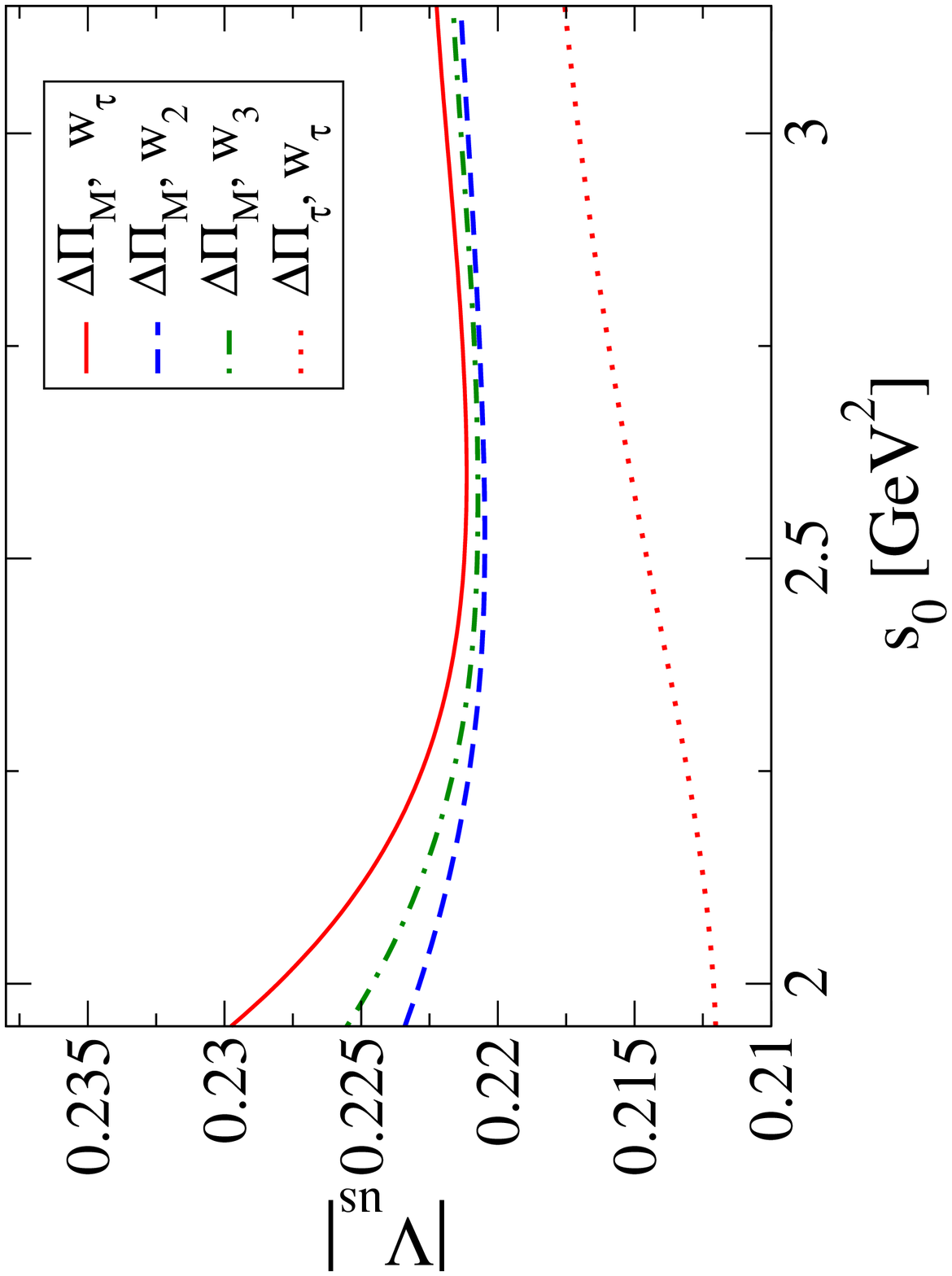}
  \end{minipage}}}\ \ 

\caption{$\vert V_{us}\vert$ vs. $s_0$ for (i) Left panel: the 
$w_\tau$ $us$ V+A FESR, using the three $D=2$ OPE prescriptions, and
(ii) Right panel: a selection of EM-$\tau$ FESRs.}
\label{usvpaandemtaufesrs}
\end{figure}

Slow convergence of the integrated $D=2$ OPE series and possible
missing higher multiplicity $us$ spectral strength could both account
for the $s_0$-instability of the FB $w_\tau$ FESR results. 
The latter possibility can be tested using FESRs for
$\Pi^{(0+1)}_{V+A;us}$. For $w(s)\geq 0$ and $s_0$ large enough 
that the region of missing strength overlaps the range of the $us$ spectral 
integral, $\vert V_{us}\vert$ should come out low, while for $s_0$ low 
enough to exclude such overlap, $\vert V_{us}\vert$ should rise back to its
true value. Two new OPE terms enter these FESRs: 
the $D=0$ contribution (known to 5-loops~\cite{bckd0ope08}) 
and a $D=4$ gluon condensate contribution.
Excellent agreement between the world average $\alpha_s$ value
and that obtained from $ud$, $J=0+1$ V, A and V+A 
FESRs~\cite{my08} shows these ingredients can be reliably evaluated.
Results for $\vert V_{us}\vert$ versus $s_0$, 
for $w=w_\tau$, are shown
in the left panel of Fig.~\ref{usvpaandemtaufesrs}.
Results for the three $D=2$ prescriptions 
agree with those of the corresponding FB $w_\tau$ FESR treatment.
The $s_0$-dependence of 
$\vert V_{us}\vert$ for the two CIPT prescriptions, however, is clearly
incompatible with the assumption that the $D=2$ OPE representation is 
reliable and the FB $w_\tau$ instability is due to missing higher
multiplicity $us$ spectral strength. As for the FB $w_\tau$
FESR, the FOPT $D=2$ treatment produces improved, though not perfect,
$s_0$-stability.

The larger-than-expected $D=2$ OPE uncertainties 
of the FB $\tau$ FESRs can be reduced by considering 
FESRs for $\Delta\Pi_M= 
9\Pi_{EM} -6\Pi^{(0+1)}_{V;ud} + \Delta\Pi_\tau$~\cite{kmtauem08}.
$\Pi_{EM}$ is the electromagnetic (EM) correlator, whose spectral function 
is determined by the bare $e^+e^-\rightarrow hadrons$ cross-sections.
$\Delta\Pi_M$ is the unique FB EM-$\tau$ combination 
with the same $\Pi^{(0+1)}_{V+A;us}$ normalization as
$\Delta\Pi_\tau$ and zero $O(\alpha_s^0)$ $D=2$ coefficient.
The $O(\alpha_s^0)$ $D=4$ coefficient is also
$0$ and the remaining $D=2$ coefficients suppressed
by factors of $\sim 5-7$ relative to those of $\Delta\Pi_\tau$.
Integrated $D>4$ contributions, which are not suppressed~\cite{kmtauem08},
can be fitted to data due to their stronger $s_0$-dependence.
The strong suppression of $D=2$ and $D=4$ contributions
at the correlator level greatly reduces OPE-induced
uncertainties~\cite{kmtauem08}. 
At present, use of these FESRs is complicated by 
inconsistencies (within isospin breaking corrections) of the EM 
and $\tau$ $2\pi$ and $4\pi$ spectral data~\cite{newdavieretalgmuetc}. We
illustrate the improved $s_0$-stability of the $\Delta\Pi_M$
FESRs in the right panel of Fig.~\ref{usvpaandemtaufesrs} for
$w=w_\tau$, $w_2(y)=(1-y)^2$ and $w_3(y)=1-{\frac{3}{2}}y+{\frac{1}{2}}y^3$,
assuming the $\tau$ data to be correct for both $2\pi$ and $4\pi$. 
The $s_0=m_\tau^2$, $w_\tau$ result for $\vert V_{us}\vert$ is 
$0.2222(20)_\tau (28)_{EM}$, with only experimental errors shown. 
$\Delta\Pi_M$ FESRs, while promising for the future,
require resolution of the $\tau$ vs. EM $2\pi$ and
$4\pi$ discrepancies.

\end{document}